\documentclass[twocolumn,showpacs,preprintnumbers,aps,pra,amsmath,amssymb,floatfix]{revtex4-1}

\usepackage{graphicx}
\usepackage{dcolumn}
\usepackage{bm}
\usepackage{epsfig}
\usepackage{color}
\usepackage{natbib}

\begin{document}

\title{A double species $^{23}$Na and $^{87}$Rb Bose-Einstein condensate with tunable miscibility via an interspecies Feshbach resonance}


\author{Fudong Wang$^{1}$, Xiaoke Li$^{1}$, Dezhi Xiong$^{1,2}$, Dajun Wang$^{1,3}$}
\email{djwang@phy.cuhk.edu.hk}
 \affiliation{
$^{1}$Department of Physics, The Chinese University of Hong Kong, Hong Kong SAR, China\\
$^{2}$State Key Laboratory of Magnetic Resonance and Atomic and Molecular Physics, and Key Laboratory of Atomic Frequency Standards (KLAFS), WIMP, CAS, Wuhan, China\\
$^{3}$The Chinese University of Hong Kong ShenZhen Research Institute, ShenZhen, China}

\date{\today}

\begin{abstract}

We have realized a dual-species Bose-Einstein condensate (BEC) of $^{23}$Na and $^{87}$Rb atoms and observed their immiscibility. Because of the favorable background intra- and inter-species scattering lengths, stable condensates can be obtained via efficient evaporative cooling and sympathetic cooling without the need for fine tuning of the interactions. Our system thus provides a clean platform for studying inter-species interactions driven effects in superfluid mixtures. With a Feshbach resonance, we have successfully created double BECs with largely tunable inter-species interactions and studied the miscible-immiscible phase transition.


\end{abstract}

\maketitle


\section{\label{sec:level1} Introduction}

Degenerate quantum gas mixtures have been a subject of intensive study in recent years. First realized with two hyperfine states of $^{87}$Rb atoms \cite{Myatt97}, such mixtures were soon extended to Bose-Bose \cite{Modugno02,Thalhammer08,Papp08,Lercher11,McCarron11}, Bose-Fermi \cite{Hadzibabic02,*Roati02} and Fermi-Fermi \cite{Taglieber08} combinations of two different atoms. Many important properties of these mixtures, such as miscibility, are determined by the relation between the inter- and intra-species interactions \cite{Ho96,Esry97,Pu98a,Timmermans98} which can be controlled with Feshbach resonances \cite{Papp08, Tojo10}.     

The marriage of heteronuclear quantum gas mixture with Feshbach resonance is also a promising gateway toward a quantum degenerate gas of ground-state ultracold polar molecules, which can introduce the long-range, anisotropic dipolar interaction to the playground \cite{Carr09,Baranov12}. This has been successfully demonstrated with $^{40}$K$^{87}$Rb, where weakly-bound Feshbach molecules were created first and then transferred to the lowest energy level with a stimulated Raman process \cite{Ni08}. To address KRb molecule's large chemical reaction induced loss \cite{Ospelkaus10,Ni10} and relatively small electric dipole moment \cite{Aymar05}, other mixtures \cite{Lercher11,McCarron11,Takekoshi12,Park12,Wu12} are now being studied intensively. Very recently, ground-state RbCs~\cite{Takekoshi2014,*Molony2014} and NaK molecules~\cite{Park2015} were successfully produced. 

$^{23}$Na and $^{87}$Rb are the first two atomic species to be Bose condensed and their mixture is among the first to be investigated theoretically \cite{Ho96,Pu98a,Pu98b}. The Na-Rb system is a nice candidate for pursuing ultracold polar molecule as the NaRb molecule has a large permanent electric dipole moment \cite{Aymar05} and is stable against two-body chemical reactions \cite{Zuchowski10}. In addition, compared with several other popular Bose-Bose mixtures, the Na-Rb system has the advantage that each individual species' background interaction can support its stable condensate. Feshbach resonances can thus be applied to tune the inter-species interaction at will for a range of interesting studies, such as non-equilibrium dynamics with quenched miscibility \cite{Sabbatini11,Hofmann2014,Bisset2015} and binary spinor condensates \cite{Shi10,Xuzhifang10,Xuzhifang12}. 

In this paper, we describe our work on the production of a double BEC of $^{23}$Na and $^{87}$Rb in a simple hybrid quadrupole + optical dipole trap setup. While the two condensates are immiscible with their background interactions, with an inter-species Feshbach resonance, we demonstrate miscibility tuning over a large range. These experimental results are well accounted for by numerical solutions of the coupled Gross-Pitaevskii (G-P) equations with interaction parameters derived from our previous Feshbach resonance work \cite{Wangfudong13}.  

The rest of this paper is organized as follows. In Sec.~\ref{sec:theory}, we present the coupled G-P equations and apply them to the $^{23}$Na and $^{87}$Rb double BEC system. This is followed by the description of how we prepare the double BEC in Sec.~\ref{sec:exp}. In Sec.~\ref{sec:results}, we show the main experimental results and their comparison with numerical simulations. Sec.~\ref{sec:conclusions} concludes the article.

\section{\label{sec:theory} Theory }

In the mean-field formalism, the ground state wavefunctions $\psi_i$ ($i$ = 1 for Na and 2 for Rb) of two overlapped BECs with particle numbers $N_i$ can be obtained from a pair of coupled time-independent G-P equations
\begin{align}
(-\frac{\hbar^2 \nabla^2}{2m_1} + V_1 + N_1 g_{11}|\psi_1|^2 + N_2 g_{12}|\psi_2|^2 )\psi_1 = \mu_1 \psi_1, \\
(-\frac{\hbar^2 \nabla^2}{2m_2} + V_2 + N_2 g_{22}|\psi_2|^2 + N_1 g_{12}|\psi_1|^2 )\psi_2 =\mu_2 \psi_2.
\label{eq:cgp}
\end{align}
Here $m_i$, $V_i$, and $\mu_i$ are the mass, external trap potential, and chemical potential for the $i$th atomic species, respectively. The two-body intra- and inter-species interactions are determined by the corresponding scattering length $a_{ij}$ via the interaction constant $g_{ij} = 2 \pi \hbar^2 a_{ij}/\mathcal{M}_{ij}$, with $\mathcal{M}_{ij}$ the reduced mass. The miscibility is determined by the relation between $g_{\mathrm{12}}$, $g_{\mathrm{11}}$ and $g_{\mathrm{22}}$. In the Thomas-Fermi limit with the kinetic energy terms ignored, phase separation happens when $g_{\mathrm{12}}> \sqrt{g_{\mathrm{11}} g_{\mathrm{22}}}$.

The two condensates are trapped with the same 1070 nm optical dipole trap (ODT). Due to the very different polarizabilities and masses between Na and Rb, their trap frequencies $\omega_i$ are different. Contributions from gravity are thus included to take into account the differential sagging caused center-to-center separation between the two condensates. The overall trap potential can be expressed as $V_i = \frac{1}{2}m_i\omega^{2}_{i}r^2+m_i g z$, with gravity along the $z$ direction.

The scattering lengths for $^{23}$Na ($a_\mathrm{11} = 54.5~a_\mathrm{B}$) \cite{Knoop11} and $^{87}$Rb ($a_\mathrm{22} = 100.4~a_\mathrm{B}$) \cite{Marte02} are well known. Here $a_\mathrm{B}$ is the Bohr radius. With the recently measured $a_\mathrm{12} = 73~a_\mathrm{B}$ \cite{Wangfudong13}, we find that $g_{\mathrm{12}}^2/g_{\mathrm{11}}g_{\mathrm{22}} > 1$, which indicates that the inter-species interaction is strong and double $^{23}$Na and $^{87}$Rb BEC with their background scattering lengths should be immiscible. The three positive but moderate magnitude scattering lengths support stable condensates. They also ensures fast elastic collisions for efficient evaporative and sympathetic cooling without causing much inelastic three-body losses, and are thus quite ideal for double BEC production. This is a big advantage over several other systems \cite{Papp08,Lercher11,McCarron11}, in which Feshbach resonances had to be employed to control the scattering length of one of the species in order to achieve the double BEC. Tuning inter-species interactions simultaneously are thus difficult for those cases. Here, we can use inter-species Feshbach resonances at will for investigating inter-species interaction driven effects. 

\begin{figure}[hbtp]
\includegraphics[width=0.85 \linewidth]{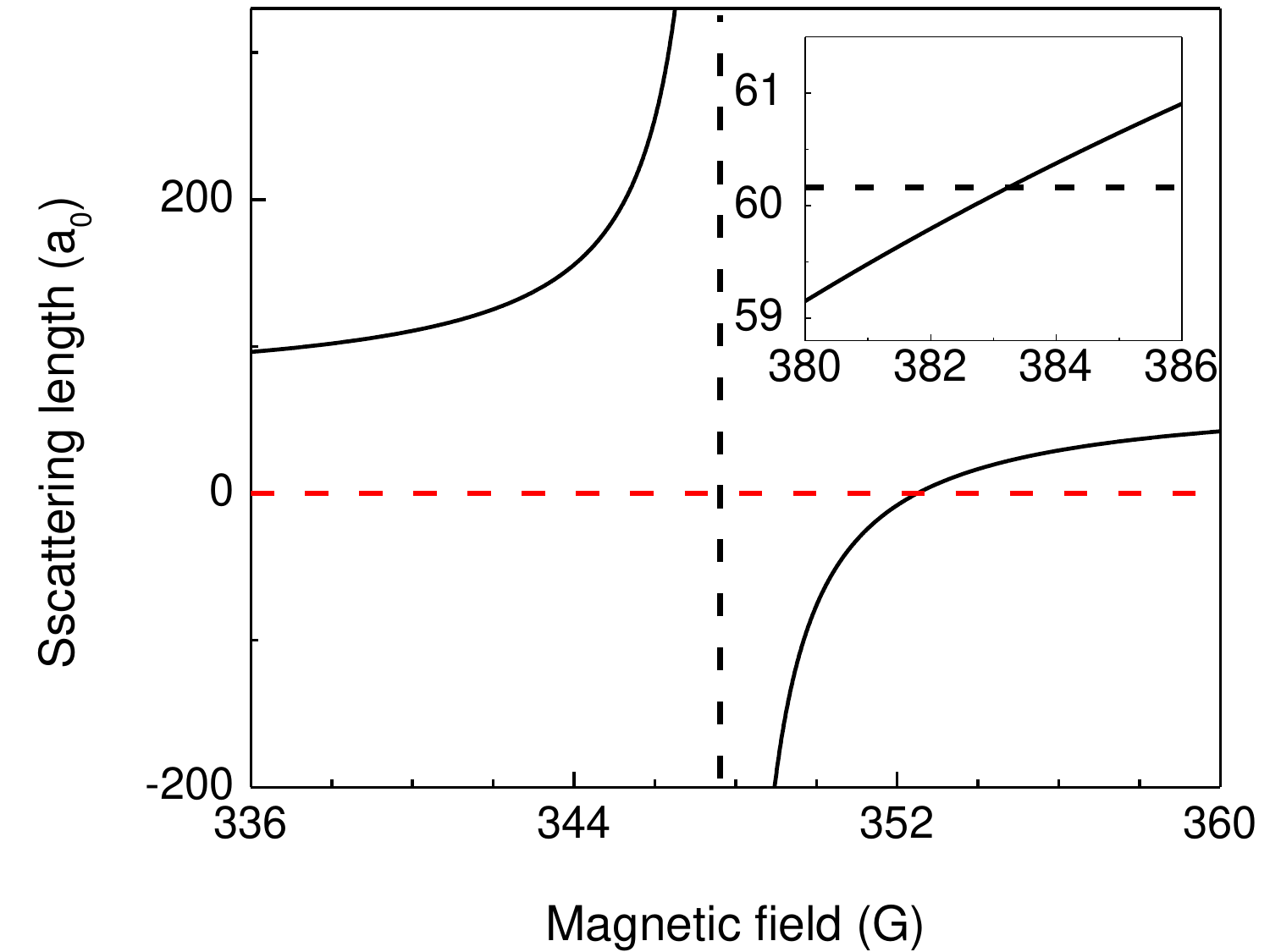}
\caption{\label{fig:Feshbach}(color online). 
Tunability of the s-wave interactions between Na and Rb atoms prepared in their lowest hyperfine states with a magnetic Feshbach resonance. The vertical dashed line indicates the resonant field at 347.8 G. Inset: near 383.2 G, $a^c_{12}$ = 60.2$a_B$ and the double BEC undergoes a miscible to immiscible transition.}
\end{figure}

Near a Feshbach resonance, $a_{\mathrm{12}}$ varies as a function of the $B$ field as
\begin{equation}
\label{eq:resonance}
a_{\mathrm{12}}=a_{\mathrm{bg}}(1+\frac{\Delta}{B-B_{0}}+...),
\end{equation}
where $a_{\mathrm{bg}}$ is the background scattering length in the vicinity of the resonant field $B_0$, and $\Delta$ is the resonance width defined as the field difference between the zero crossing point and the resonance. To make fine tuning of interaction with realistic $B$ field resolution and stability, it is desirable to have a large $\Delta$. In principle, all resonances in the same channel should be included here for an accurate calculation of $a_{12}$.

As illustrated in Fig.\ref{fig:Feshbach}, the 347.8 G $s$-wave resonance between $^{23}$Na and $^{87}$Rb atoms \cite{Wangfudong13} in their lowest-energy states has a width $\Delta = $ -4.9 G. Thus $a_{\mathrm{12}}$ and the interspecies interactions can already be changed over a large range without going too close to the resonance where three-body loss becomes large. Near the zero crossing point and the nominal miscible-immiscible transition point, $a_{\mathrm{12}}$ varies with very small slopes. It is thus quite feasible to experimentally control $a_{\mathrm{12}}$ with a resolution $<<$1 $a_\mathrm{B}$ around these magnetic field regions for studying non-interacting mixtures and miscibility related problems with high precision.

\section{\label{sec:exp} Double BEC production in a hybrid trap }

\begin{figure}[hbtp]
\includegraphics[width=0.9\linewidth]{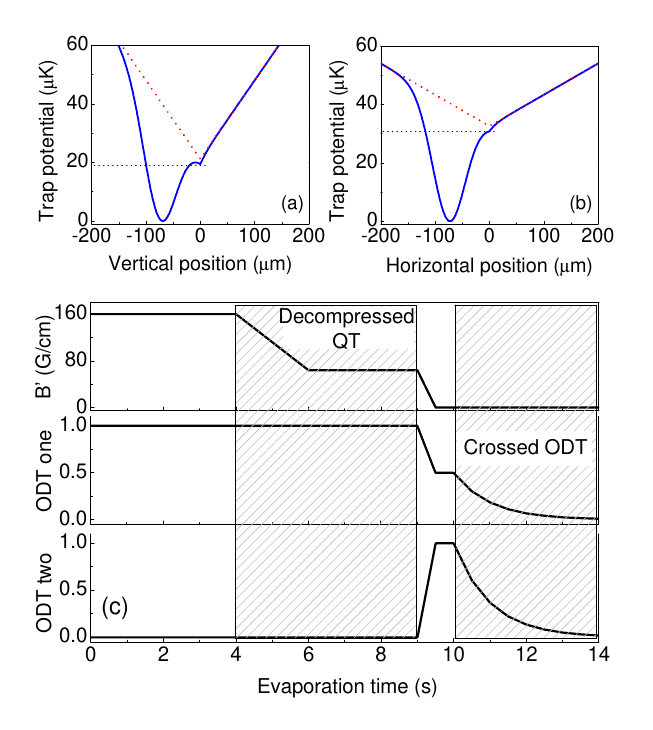}
\caption{\label{fig:potential}(color online). Comparison of the original (a) and the modified (b) hybrid trapping potentials (blue solid lines) after decompression. The ODT displacements from the QT center are 75 $\mu$m in both cases. The potential walls from the potential minimum to the center of the QT depicted by the horizontal dashed lines are 19 $\mu$K and 31 $\mu$K for configurations (a) and (b), respectively. The QT potentials are shown in red dashed lines for comparison. The evolution of the QT and ODT powers during the evaporation sequence in the hybrid trap are shown schematically in (c). }
\end{figure}

Our single vacuum chamber apparatus has been descried previously\cite{Xiong13,Wangfudong13}. Here we present the major modifications only. The most important one is on the configuration of the hybrid trap based on a better understanding of its operation principle. Different from the original implementation \cite{Lin09}, which had the ODT focus displaced vertically, the displacement of our ODT focus is along the horizontal (radial) direction of the magnetic quadrupole trap (QT). As illustrated in Fig.~\ref{fig:potential}(a) and (b) for $^{23}$Na atoms, with the same ODT and QT parameters, our configuration provides a higher potential wall from the potential minimum to the QT center. This potential wall is essential in preventing Majorana loss from happening. While this modification makes little difference during the initial stage of the evaporation, it becomes more important when the clouds become colder where Majorana loss is severe. 

Another important change we put forward is on the QT. As shown in Fig.~\ref{fig:potential}(c), now we decompress it by ramping the gradient from 160 G/cm down to 64 G/cm in 2 s after the microwave frequency is scanned to 6822 MHz. The atom clouds are cooled adiabatically during this process, meanwhile the aforementioned potential wall is increased by 50$\%$, which suppresses Majorana loss further. With these two improvements we have been able to produce a $^{23}$Na cloud cold enough for an efficient crossed ODT loading. The microwave frequency scan continues during the decompression and keeps going on in the decompressed QT.

As our Na MOT contains only $5\times10^6$ $^{23}$ atoms, sympathetic cooling is necessary. Thanks to the favorable $a_{\mathrm{12}}$ and the much larger number of Rb atoms ($3\times10^8$ atoms), sympathetic cooling is still rather efficient even in the decompressed QT. This is evident by the fact that the same microwave evaporation protocol optimized for Rb atoms only can be used for the two species evaporation without further tuning.

Evaporation in the decompressed QT stops at 6833.98 MHz where only $4.7\times10^4$ $^{87}$Rb atoms are left with $4.2\times10^5$ $^{23}$Na atoms. At this point, the $^{23}$Na($^{87}$Rb) temperature is 2.5 $\mu$K (2.3 $\mu$K) and the calculated phase-space density (PSD) is 0.058(0.006) for $^{23}$Na($^{87}$Rb). We note that sympathetic cooling has already stopped when the coolant $^{87}$Rb atoms are less than $^{23}$Na atoms at about 6833.75 MHz. The further removal of $^{87}$Rb atoms is for the benefit of the pure ODT loading and the following ODT evaporation. In the same 1070-nm ODT, the trap depth for $^{23}$Na atom $U_{\mathrm{Na}}$ is $\sim$3 times less than than  $U_{\mathrm{Rb}}$ of $^{87}$Rb atoms. Thus when we ramp down the ODT power for evaporation, $^{23}$Na always evaporates faster and it becomes the coolant while $^{87}$Rb is sympathetically cooled. With an excessive amount of $^{87}$Rb atoms as the thermal load, we have found that it is impossible to create the double BEC. 

The atoms are then loaded into the crossed ODT by simultaneously lowering down the magnetic field gradient to zero and ramping up a second laser beam in 500 ms, as illustrated in Fig.~\ref{fig:potential}(c) in between the two shaded regions. Both ODT beams are of 65 $\mu$m beam waist and they intercept each other with an angle of 62$^{\circ}$. A 16 G homogeneous magnetic field is applied at the end of the loading to provide a quantization axis for the spin polarized atoms. Typically, about 70$\%$ of the $^{23}$Na atoms and almost all $^{87}$Rb atoms can be loaded into the crossed ODT. The PSDs are improved to 0.19(0.02) for $^{23}$Na($^{87}$Rb), due to the trap geometry deformation and also continuous evaporation during this loading procedure\cite{Lin09}. In the crossed ODT, the trap frequencies $\omega_{\mathrm{Na}} : \omega_{\mathrm{Rb}} = \sqrt{U_{\mathrm{Na}}/m_{\mathrm{Na}}} : \sqrt{U_{\mathrm{Rb}}/m_{\mathrm{Rb}}} \approx$ 1.1. Assuming thermal equilibrium, the sizes of the two clouds $\sigma_{\mathrm{Na}} : \sigma_{\mathrm{Rb}} \approx$ 1.8. Thus using $^{23}$Na as the coolant is advantageous for sympathetic cooling because $^{87}$Rb atoms are always immersed in the $^{23}$Na bath. Indeed, a high sympathetic cooling efficiency of 3.4 has been observed for $^{87}$Rb.  

Continuous evaporation in the crossed ODT leads to a bimodal distribution in $^{23}$Na first which signifies its BEC phase transition, while that for $^{87}$Rb always lags behind. We attribute this to $^{87}$Rb atom's lower transition temperature $k T_c \approx 0.94\hbar \omega N^{1/3}$ because of its smaller atom number $N$ and lower trap frequency $\omega$. Here $k$ is the Boltzmann's constant. In the end, we are able to produce a quasi-pure double BEC with $2.4\times10^4$ ($1.0\times10^4$) $^{23}$Na($^{87}$Rb) atoms. The whole ODT evaporation lasts for 4.3 s. The final trap frequencies are measured with parametric resonances to be 2$\pi \times$(124,143,74) and 2$\pi \times$(112,129,66)Hz for $^{23}$Na and $^{87}$Rb, respectively. We can also produce a single species $^{23}$Na BEC with $1.5\times10^5$ atoms, if we remove all $^{87}$Rb atoms at the microwave evaporation step.

\section{\label{sec:results} Results and discussions }

\begin{figure}[hbtp]
\includegraphics[width=0.85 \linewidth]{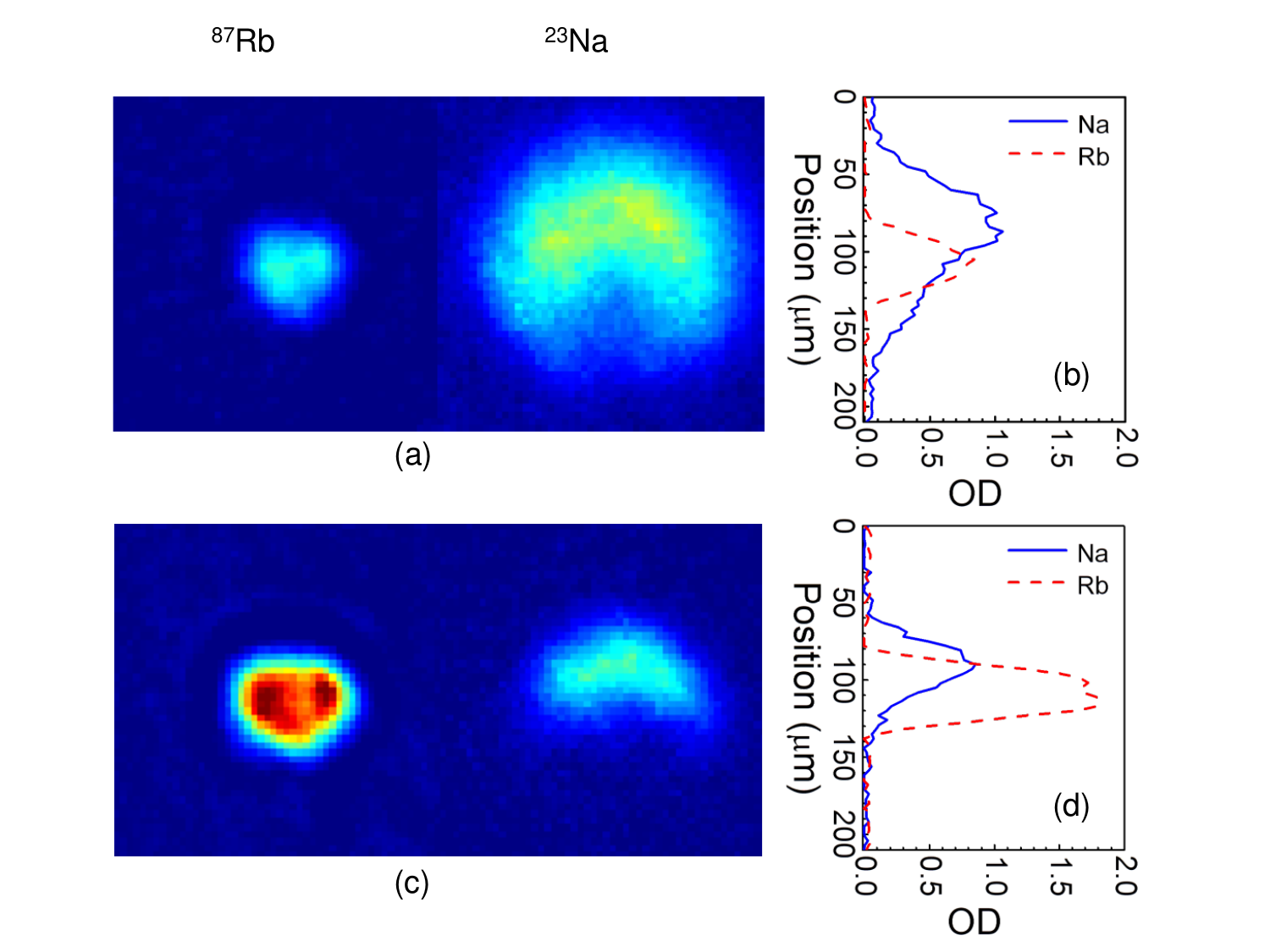}
\caption{\label{fig:BEC}(color online). 
Absorption images after 10 ms TOF which show immiscibility of the double BEC with their background interaction. (a) and (c) are pairs of a $^{87}$Rb and a $^{23}$Na condensates with different atom number ratios. Field of view: 210 $\mu$m $\times$ 210 $\mu$m. (b) and (d) are the center cross sections along the vertical direction for images in (a) and (c), respectively. }
\end{figure}

Immiscibility of the two simultaneously condensed samples shows up clearly in the time-of-flight (TOF) absorption images taken from the horizontal direction. Fig.~\ref{fig:BEC} demonstrates this with two very different atom number ratios. Images in Fig.~\ref{fig:BEC}(a) contain $3.5\times10^3$($3.2\times10^4$) $^{87}$Rb($^{23}$Na) atoms, while those of Fig.~\ref{fig:BEC}(c) have $8.5\times10^3$($1.3\times10^4$) $^{87}$Rb($^{23}$Na) atoms. It is apparent that the presence of $^{87}$Rb will always repel $^{23}$Na atoms away, in agreement with the inter- and intra-species interaction ratios.

We note that the two clouds are not concentric to each other because of the differential gravitational sag in the crossed ODT. For the typical final vertical trap frequencies, this effect makes the $^{87}$Rb cloud center-of-mass (COM) locates 2.8 $\mu$m below that of $^{23}$Na. Taking the samples in the upper images as examples, the calculated single species Thomas-Fermi radius is $5.5~\mu$m for $^{23}$Na and $2.6~\mu$m for $^{87}$Rb. Thus the $^{87}$Rb cloud can only overlap and repel the lower part of the $^{23}$Na one, consistent with the crescent shape $^{23}$Na images and the optical density (OD) cross sections in Figs.~\ref{fig:BEC}(b) and (d).

\begin{figure}[hbtp]
\includegraphics[width=0.85 \linewidth]{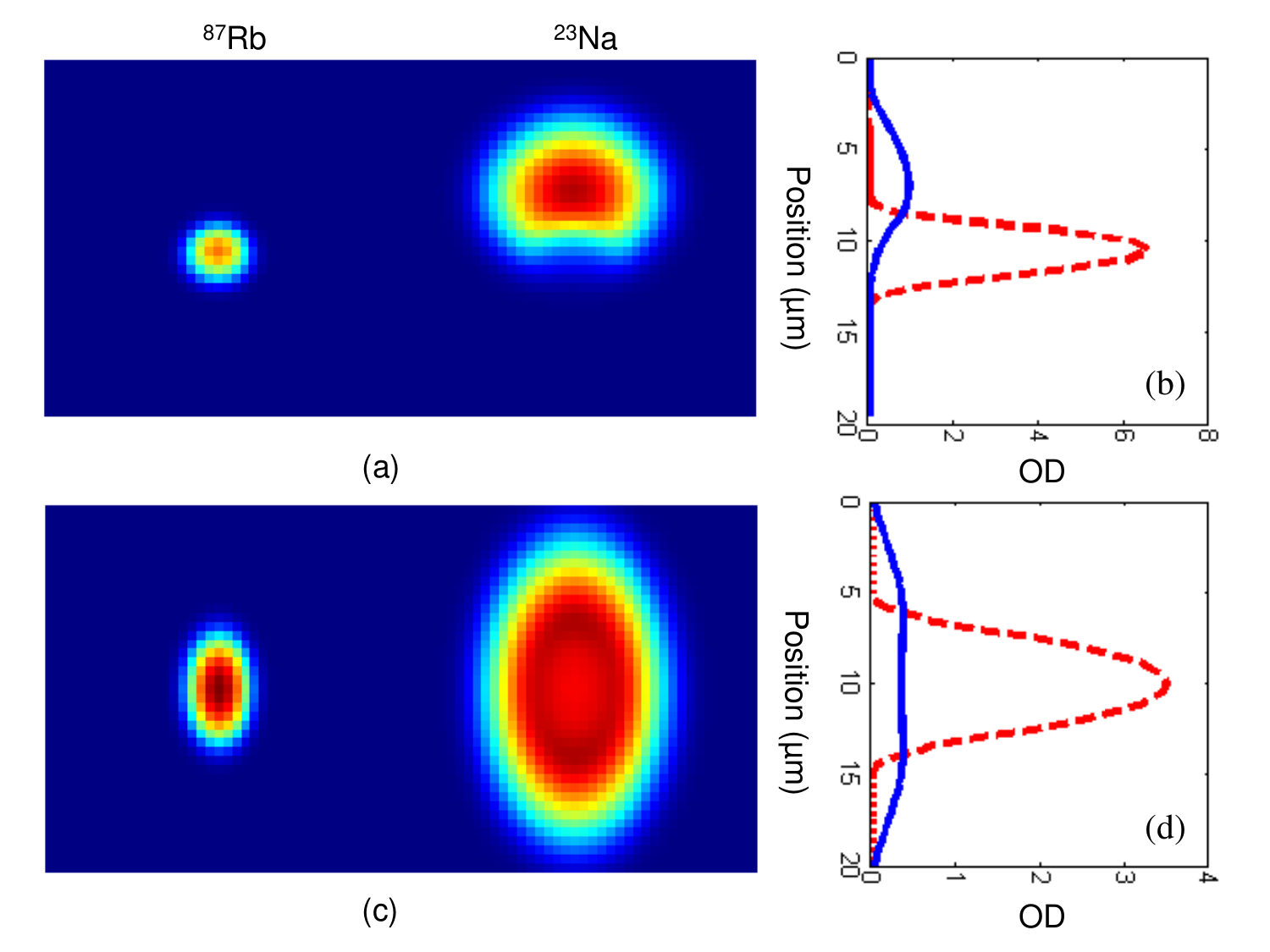}
\caption{\label{fig:BECThe}(color online). 
Simulated in-trap column density profiles with the same experimental conditions as those in Fig.~\ref{fig:BEC}(a). The probing directions are horizontal for (a) and vertical for (c). Field of view: 20 $\mu$m $\times$ 20 $\mu$m. The contrasts of the Na images are adjusted to enhance the visibility. (b) and (d) are the corresponding cross sections with the blue solid curve for Na and the red dashed curve for Rb.}
\end{figure}

This qualitative understanding is supported by numerical simulations. As depicted in Fig.~\ref{fig:BECThe}(a) are in-trap density profiles obtained by numerically solving the coupled G-P equations with the corresponding experimental conditions for producing the images in Fig.~\ref{fig:BEC}(a). The overall experimentally observed density patterns are qualitatively captured by the simulation. One of the main features, the notch at the lower part of the Na cloud due to the repulsive Na-Rb interactions, is well reproduced. From the vertical cross section in Fig.~\ref{fig:BECThe}(b), the displacement between the center of the two clouds is determined to be 3.5 $\mu$m. The additional 0.7 $\mu$m compared with the differential gravitational sag reflects repulsive nature of the inter-species interaction. We note that the experimentally measured vertical displacement from Fig.~\ref{fig:BEC}(b) are much larger due to the expansion\cite{Esry97}.

Limited by optical access, our setup does not have the capability of taking absorption images from the vertical direction. Judging from symmetry and the horizontal images, we can reasonably conjecture that the two images should be concentric with each other in that direction, with the smaller size $^{87}$Rb cloud totally wrapped around by $^{23}$Na atoms. This is supported by the simulated density patterns and cross sections in Fig.~\ref{fig:BECThe}(c) and (d), respectively. As a result of the interspecies interaction, the $^{23}$Na BEC density becomes flat-topped. In future investigations, besides improving the observation capability, it may also worth to use another laser wavelength for the ODT to eliminate the differential gravitational sag so that the two clouds can be concentric in all 3 directions. Such a ``magic wavelength'' has been calculated to be 946.66 nm for $^{23}$Na and $^{87}$Rb \cite{Safronova06}. 

\begin{figure*}
\centering\includegraphics[width=0.9\linewidth]{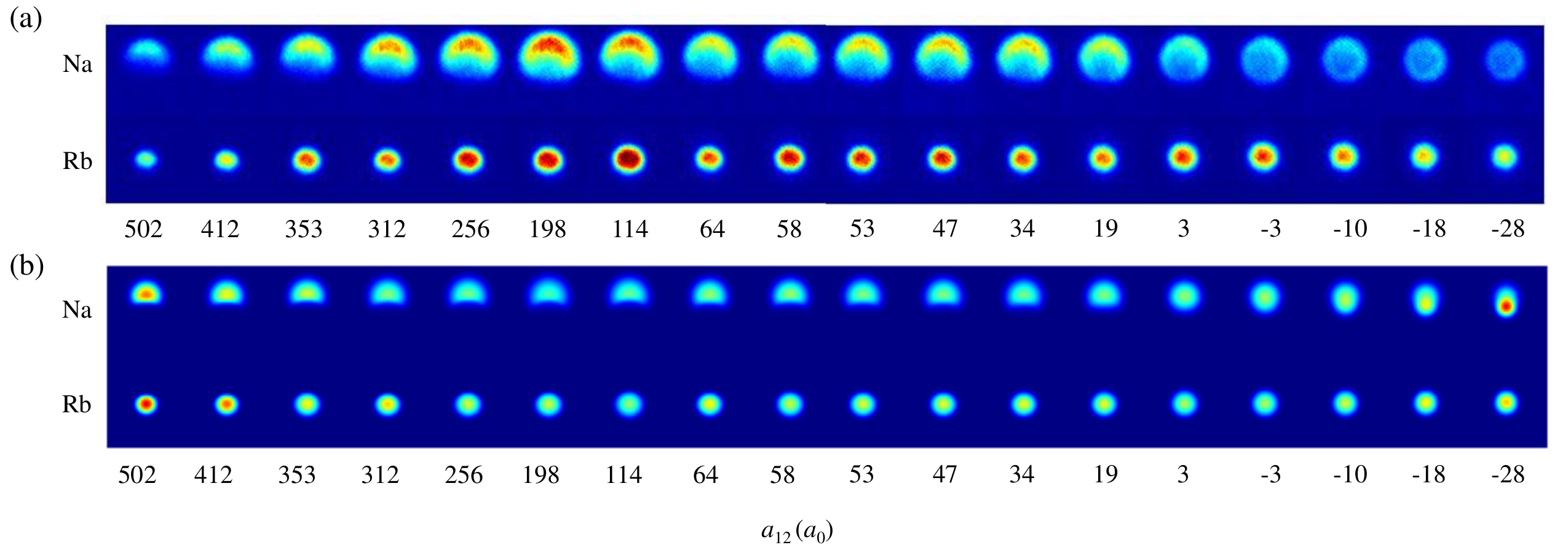}
\caption{(color online) (a) Column density profiles of the double BEC near the 347.8 G Feshbach resonance measured after 13 ms expansion. The inter-species scattering lengths $a_{12}$ are calculated from Eq.~(\ref{eq:resonance}) for each field values. Note that atom numbers are not the same for all interaction strengths due to three-body losses. (b) Simulated in-trap column density profiles for the double BEC near the 347.8 G Feshbach resonance. The simulation conditions for each image are the same as the corresponding experimental ones in (a).}
\label{fig:tunable}
\end{figure*}

Since fine tuning of interaction is not needed at the double BEC creation stage, Feshbach resonance can be used solely for studying interspecies interaction induced effects. Here we study the miscible-immiscible phase transition with the 347.8 G inter-species resonance. We first prepare both atoms in their $\left|1, 1\right\rangle$ hyperfine levels with an adiabatic rapid passage after loading the atoms into the crossed ODT. The magnetic field is then brought up to a range of values near the 347.8 G Feshbach resonance while ODT evaporation is performed. Once the final ODT power is reached, both the ODT and the magnetic field are turned off abruptly and simultaneously after 200 ms holding. The atoms are allowed to expand for 13 ms before absorption images are taken.

%
%

As shown in Fig.~\ref{fig:tunable}(a), the miscibility can be changed dramatically by tuning $a_{\rm 12}$ from -28 $a_{\rm B}$ to 502 $a_{\rm B}$. At negative $a_{\rm 12}$, the mutual attraction pulls the two clouds together and greatly increases their overlap. For more negative $a_{\rm 12}$, we observe inter-species attraction induced collapsing (images not shown). As $a_{\rm 12}$ increases, the center to center separation between the two condensates becomes larger. The remaining atom numbers also become less due to increasing three-body losses. While for even larger positive $a_{\rm 12}$, these losses and the accompanying heating become too severe to perform the measurement. The observed miscibility character agrees qualitatively with the simulated ones, as presented in Fig.~\ref{fig:tunable}(b).

\begin{figure}
\includegraphics[width=0.85 \linewidth]{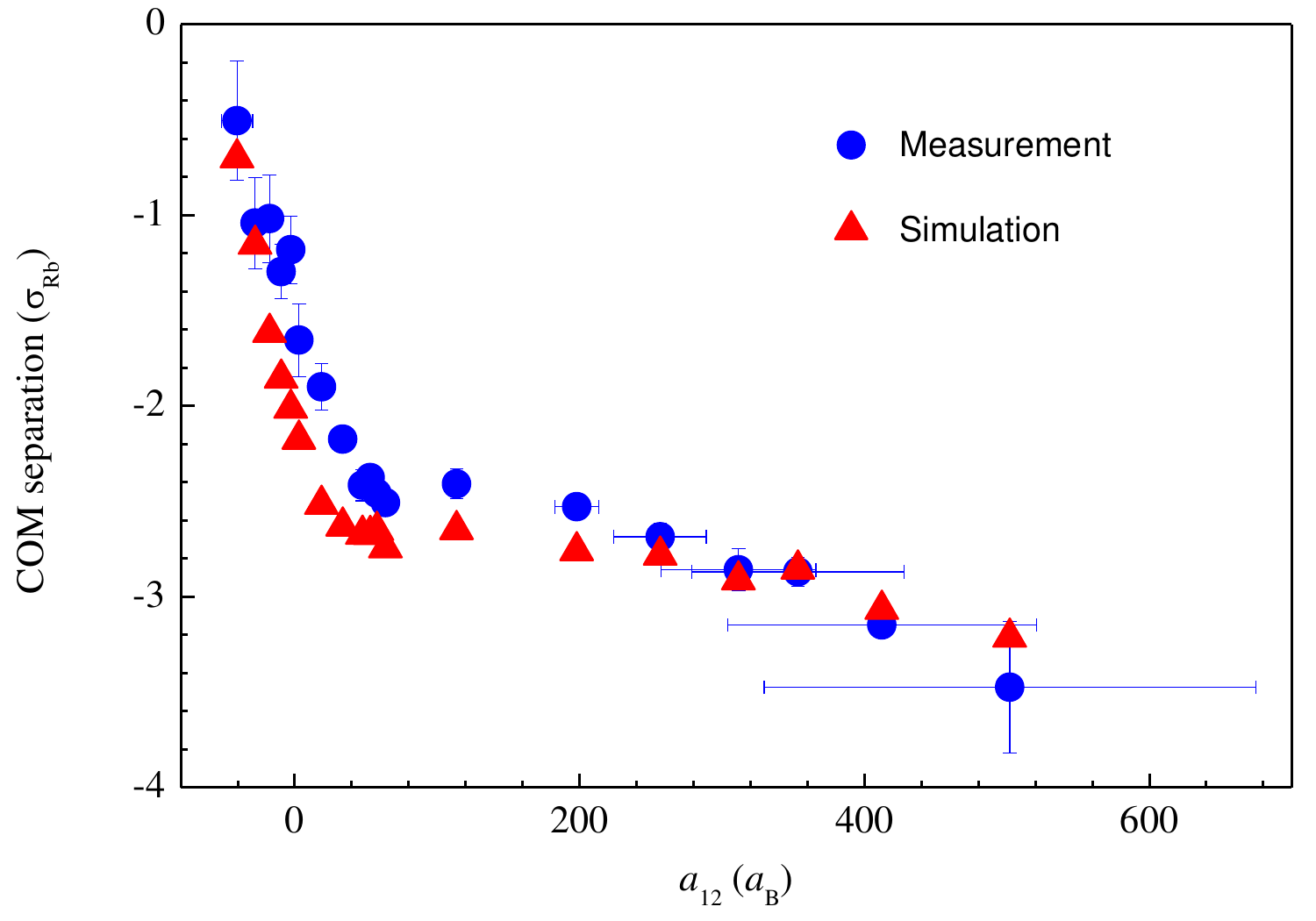}
\caption{\label{fig:com}(color online). Measured and simulated vertical COM separation of the Na and Rb clouds vs. $a_{12}$. The data are extracted from images in Fig.~\ref{fig:tunable}(a) and (b), respectively. All data are normalized to the Rb condensate sizes under each corresponding $a_{12}$. Error bars in the measured separations represent one standard deviation of statistics from typically 5 repeated shots, while those of $a_{12}$ are from $B$ field resolution.  
}
\end{figure}
    
At the critical inter-species scattering length $a_{\rm 12}^c \approx 60 a_{\rm B}$, we cannot identify obvious signatures of miscible-immiscible transition from the images. To characterize the miscibility vs. $a_{12}$ more quantitatively, we extract the vertical COM separation between each pair of clouds from Fig.~\ref{fig:tunable} (a). As illustrated in Fig.~\ref{fig:com} are the separations normalized to the Rb condensate size. The normalization is necessary to partially cancel out the atom numbers variation effect. There is a noticeable kink near $a_{\rm 12}^c$, below which the separation decreases with a steep slope. From column density profiles in Fig.~\ref{fig:tunable}(b), the simulated COM separation vs. $a_{12}$ is also plotted in Fig.~\ref{fig:com} in the same manner. The simulation agrees qualitatively with our measurement, especially the transition point is clearly reproduced.

\section{\label{sec:conclusions} Conclusions }

To conclude, we have successfully created a double BEC of $^{87}$Rb and $^{23}$Na atoms with widely tunable inter-species interactions. The miscible-immiscible phase transition is observed, but the transition is not very sharp. This indicates that the kinetic energy terms are important in double BECs, especially for the relatively small atom number situation here. 

In future works, it will be interesting to study the same phenomena with much larger condensates when the Thomas-Fermi approximation is more valid. In-situ detection method may also be developed to extract more information necessary for investigating dynamics during the miscible-immiscible phase transition~\cite{Hofmann2014,Bisset2015}. The capability of creating miscible double BEC also makes it possible to use condensates as a starting point for magneto-association to achieve the highest molecule conversion efficiency \cite{Hodby05}. 


We acknowledge financial supported from Hong Kong RGC (Grant Nos. CUHK403111, CUHK404712), and the National Basic Research Program of China (973 Program) under Grant No. 2014CB921402. D.Z.X is partially supported by the National Natural Science Foundation of China (Grant No. 11104322).


%

\end{document}